\documentclass[preprint,showpacs,pre,10pt,superscriptaddress]{revtex4-1}
\usepackage{amsfonts}
\usepackage{amssymb}
\usepackage{amsmath}
\usepackage{graphicx}

\begin{document}
\title{Quantum Ising model in transverse and longitudinal fields: chaotic wave functions }
\author{Y. Y. Atas }
\affiliation{The University of Queensland, School of Mathematics and Physics, Brisbane, Queensland 4072, Australia}
\author{E. Bogomolny}
\affiliation{Univ. Paris-Sud, CNRS, LPTMS, UMR8626, F-91405, Orsay, France}

\date{\today}
\begin{abstract}
The construction of  a statistical model  for eigenfunctions of the Ising model in transverse and longitudinal fields is discussed in detail for the chaotic case. When the number of spins is large, each wave function coefficient has the Gaussian distribution with zero mean and the variance calculated from the first two moments of the Hamiltonian. The main part of the paper is devoted to the discussion of different corrections to the asymptotic result. One type of corrections is related with higher order moments of the Hamiltonian and can be taken into account by Gibbs-like formulae. Another corrections are due to  symmetry contributions which manifest as different numbers of non-zero real and complex coefficients. Statistical model with these corrections included agrees well with numerical calculations of wave function moments.
      
\end{abstract}
\maketitle

\section{Introduction}

Direct numerical calculations of non-integrable quantum many-body problems become exponentially  difficult for large number of particles. The developing of approximate statistical description of such models has been and remain thus of great importance. 

For macroscopic bodies the number of interacting particles, $N$, is so large that only the thermodynamic limit $N\to\infty$ is meaningful. In this case the behaviour of a small sub-system is described by its reduced density matrix constructed from knowledge of eigenvalues, $E_n$, and eigenfunctions, $\psi(n)$, when the interaction with outside particles is switched off. According to general properties of quantum thermodynamics (see e.g. \cite{landau})  the mean value of an observable $\hat{A}$ of the sub-system is calculated as follows
\begin{equation}
\lim_{T\to \infty}\frac{1}{T}\int_0^T \langle \Psi(t)|\hat{A}|\Psi(t)\rangle \mathrm{d}t= \frac{1}{Z}\sum_{n}\mathrm{e}^{-\beta E_n} \langle \psi(n) |\hat{A}|\psi(n)\rangle,\qquad Z=\sum_{n}\mathrm{e}^{-\beta E_n}  
\label{gibbs}
\end{equation} 
and the only information of the existence of a larger system is the inverse temperature $\beta$ which is determined by the mean energy of the initial state. 

This text-book approach reveals so successful that there is no doubt (at least in physical literature) that it is the correct way of calculation of the thermodynamic limit in generic systems.  The existence of powerful perturbation expansion around equilibrium states with zero and non-zero temperature \cite{green_book}, and non-equilibrium states \cite{keldysh} contributes considerably to the success of quantum thermodynamics. The important particularity of these constructions is that in any steps effectively infinite number of particles never appears explicitly. The formalism is built in such a way that only finite quantities like temperature and chemical potential give information of the existence of outside world.    

With the ever increasing power of modern computers, it has become possible to perform full numerical calculations for quantum systems containing few tens of particles. This in turn enables meaningful formulations of questions about the foundations of quantum thermodynamics, which is a work in progress in modern physics: what systems are thermalized, what are corrections to thermodynamic limit, etc.

Let $E_{\alpha}$ and $\Psi(\alpha)$ be exact eigenvalues and eigenfunctions of the full $N$ particles Hamiltonian. For finite $N$, instead of  canonical Gibbs measure \eqref{gibbs}, one has to use the microcanonical average 
\begin{equation}
\lim_{T\to \infty}\frac{1}{T}\int_0^T \langle \Psi(t)|\hat{A}|\Psi(t)\rangle\mathrm{d}t=
\frac{1}{\mathcal{N}_{E_0,\Delta E_0}}\sum_{|E_0-E_{\alpha}|<\Delta E_0}\langle \Psi(\alpha)|\hat{A} |\Psi(\alpha)\rangle 
\label{microcanonical}
\end{equation}
where $E_0$ is the mean energy of the initial state, $\Delta E_0$ is  a energy window  assumed to be small with respect to $E_0$ but much larger than the distance between nearest levels
\begin{equation}
1/\rho(E_0)\ll \Delta E_0\ll E_0
\label{delta_E}
\end{equation}
where $\rho(E)$ is the mean spectral density at energy $E$.  $\mathcal{N}_{E_0,\Delta E_0}$ is the number of states in this window. (We consider only the so-called chaotic systems where there is no integral of motion except classical ones.) 

The important difference between Eq.~\eqref{gibbs}  and Eq.~\eqref{microcanonical} is that in the former $\psi(n)$ is eigenfunctions of a sub-system but in the later $\Psi(\alpha)$ are eigenfunctions of the full Hamiltonian. 

In such approach  the central object of investigation is $N$-particles eigenfunction $\Psi(\alpha)$. There exist different scenarios  about the emergence of thermodynamic behaviour (see e.g. \cite{rigol}). One of the most accepted is the so-called  eigenstate thermalization conjecture (ETH) \cite{deutsch_1}-\cite{srednicki}  which states that relation \eqref{microcanonical} is valid for (almost) all individual eigenstates. Similar  conjecture is known in low-dimensional quantum chaos (see e.g. \cite{fishman}) where it is explicitly stated that for  a smooth classical observable $A(p,q)$ almost all diagonal matrix elements of the corresponding quantum counterpart   tend to the microcanonical semiclassical limit when $\hbar\to 0$
\begin{equation}
\langle \Psi(\alpha) |\hat{A}|\Psi(\alpha) \rangle\underset{\hbar\to 0}{\longrightarrow} \frac{1}{Z}\int \delta(E-H(p,q))A(p,q)\mathrm{d}p\,\mathrm{d}q, \qquad  Z=\int \delta(E-H(p,q)) \mathrm{d}p\,\mathrm{d}q.
\label{semiclassical}
\end{equation}
The smallness of off-diagonal matrix elements 
$\langle  \Psi_{\alpha}|\hat{A}| \Psi_{\beta}\rangle\ll\langle  \Psi_{\alpha}|\hat{A}| \Psi_{\alpha}\rangle $, when  $\beta\neq \alpha$
 also appears naturally in quantum chaos (cf. e.g. \cite{fishman_2}) as only periodic orbits give the contribution to this quantity. 

These and many other arguments strongly suggest that for systems where quantum thermodynamics can be applied, the majority of eigenfunctions are universal depending only on a few parameters. It has been even proved that in (a certain sense) typical  functions in quantum mechanics lead automatically to canonical averaging \cite{tasaki}, \cite{typicality}.  The well known example of such phenomenon is given by Berry's conjecture \cite{berry}-\cite{heller} which postulates that wave functions of low-dimensional chaotic systems are Gaussian random functions with variance determined from  semiclassical microcanonical average as in Eq.~\eqref{semiclassical}.  Similar construction has been applied for various problems in nuclear and atomic physics, many-body models etc,   \cite{rigol}, \cite{french_1}-\cite{elon}.

The purpose of this paper is to construct and carefully check a statistical model for eigenfunctions of the quantum Ising model in transverse and longitudinal fields determined by the  Hamiltonian 
\begin{equation}
\mathcal{H}=-\sum_{n=1}^N \sigma_n^{x}\sigma_{n+1}^x-\lambda  \sum_{n=1}^N \sigma_n^z-\alpha \sum_{n=1}^N\sigma_n^x \ ,
\label{ising_2}
\end{equation}
where $\sigma^{x,y,z}$ are usual Pauli matrices, and parameters $\lambda$ and $\alpha$ fix values of transverse and longitudinal fields. 

This model is a prototypical example of  one-dimensional spin chains with nearest-neighbour interactions. When $\alpha=0$ it reduces to the well known quantum Ising model in transverse field which is integrable by the Jordan-Wigner transformation \cite{j_w}, \cite{lieb}
and for $\lambda=1$ it becomes critical  and  serves as paradigmatic model of quantum critical phenomenon \cite{sachdev}. 
For non-zero $\alpha$ (and $\lambda\neq 0$) the model is considered as non-integrable but at critical value of transverse field, $\lambda=1$, and  specially fine-tuned  weak longitudinal field, $\alpha\to 0$,  it is integrable but not conformal \cite{zamolodchikov}.  

In \cite{b_a} it was shown that the ground state wave function of the Ising model (as well as of practically all one-dimensional spin-chain models) is multifractal in the initial spin basis. So it does not feet the standard thermodynamic scheme. For high excited states the situation is different.   The spectral density for this model has been discussed in \cite{a_b}. For finite  number of spins, $N$,  in the bulk of the spectrum when $E\sim \sqrt{N}$ there exist two different regimes. When the both  fields $\lambda$ and $\alpha$ are of the order of $1$ (i.e.  of order of the  hopping term), the spectral density is well approximated by a simple Gaussian function whose parameters are calculated from the knowledge of the first moments of the Hamiltonian. If $\lambda$ is small or large the spectral density of the Ising model in two fields for large but finite $N$ has many peaks well described  by a sum of Gaussian functions calculated directly from the Hamiltonian without full diagonalization.   

In parallel with two regimes of spectral density of the Ising model there are two regimes for eigenfunctions in the bulk of the spectrum.  Here we consider in detail the case when all coupling constants are of the same order. Multi-peaks case will be discussed elsewhere.  

The plan of the paper is the following.  The  construction of chaotic wave functions for the Ising model is discussed in Section~\ref{statistic}.  To check the accuracy  of such statistical model it is convenient to calculate moments of wave functions and compare them with statistical approximations. In Section~\ref{participation} it is done for the participation ratio for the Ising model in two fields. 
The main part of this Section is devoted to the calculations of different corrections to the lowest order approximation.  It is demonstrated that  corrected statistical model of eigenfunctions agrees very well with the results of direct numerical calculations. The first moments of the full Ising model Hamiltonian  are determined in Appendix~\ref{higher_moments}.     

\section{Statistical description of wave functions}\label{statistic}

The construction of chaotic wave functions for different problems follow approximately the same steps as in \cite{berry}.

An eigenfunction of the Hamiltonian \eqref{ising_2}  with  energy $E_{\alpha}$ by definition obeys
\begin{equation}
\mathcal{H} |\Psi_{\alpha}\rangle =E_{\alpha} |\Psi_{\alpha}\rangle
\end{equation}
and can be represented as  a finite series of basis set functions, $ |\vec{n}\, \rangle$
\begin{equation}
|\Psi_{\alpha}\rangle =\sum_{\vec{n}} C_{\vec{n}}(\alpha) |\vec{n}\, \rangle.
\label{expansion} 
\end{equation}
For $N$ spins-$\tfrac{1}{2}$ the full dimension of the Hilbert space is $\mathcal{N}=2^N$.  
Here    a state is indicated by symbol $\vec{n}=(n_1,n_2,\ldots,n_N)$ with the convention that  $n_k=0$ corresponds to spin at  site $k$ down  and $n_k=1$ shows that site $k$ is occupied by spin up.

All information of wave function is contained in coefficients  $C_{\vec{n}}(\alpha)$. Let us consider the collection of coefficients  $C_{\vec{n}}(\alpha)$ with fixed symbol $\vec{n}$ and with energies in a small interval 
$I=[E <E_{\alpha}<E+\Delta E]$.  The energy window $\Delta E$ is assumed to be small with respect to $E$ but much larger than the distance between nearest levels as in \eqref{delta_E}.

The principal  assumption is  that for chaotic systems coefficients $C_{\vec{n}}(\alpha)$ in such intervals are so  irregular functions of eigen-energies $E_{\alpha}$ that their exact values are irrelevant for most purposes and one has to developed statistical description of chaotic wave functions. For spin-chains models without random parameters as the one given by Eq.~\eqref{ising_2} we are not aware of strict definition of chaoticity. The usual lore is that non-integrability of  systems with large number of degrees of freedom is (almost) synonym of  chaoticity. For our purposes such notion is sufficient.

\begin{figure}
\begin{minipage}{.4\linewidth}
\begin{center}
\includegraphics[width=.95\linewidth, clip]{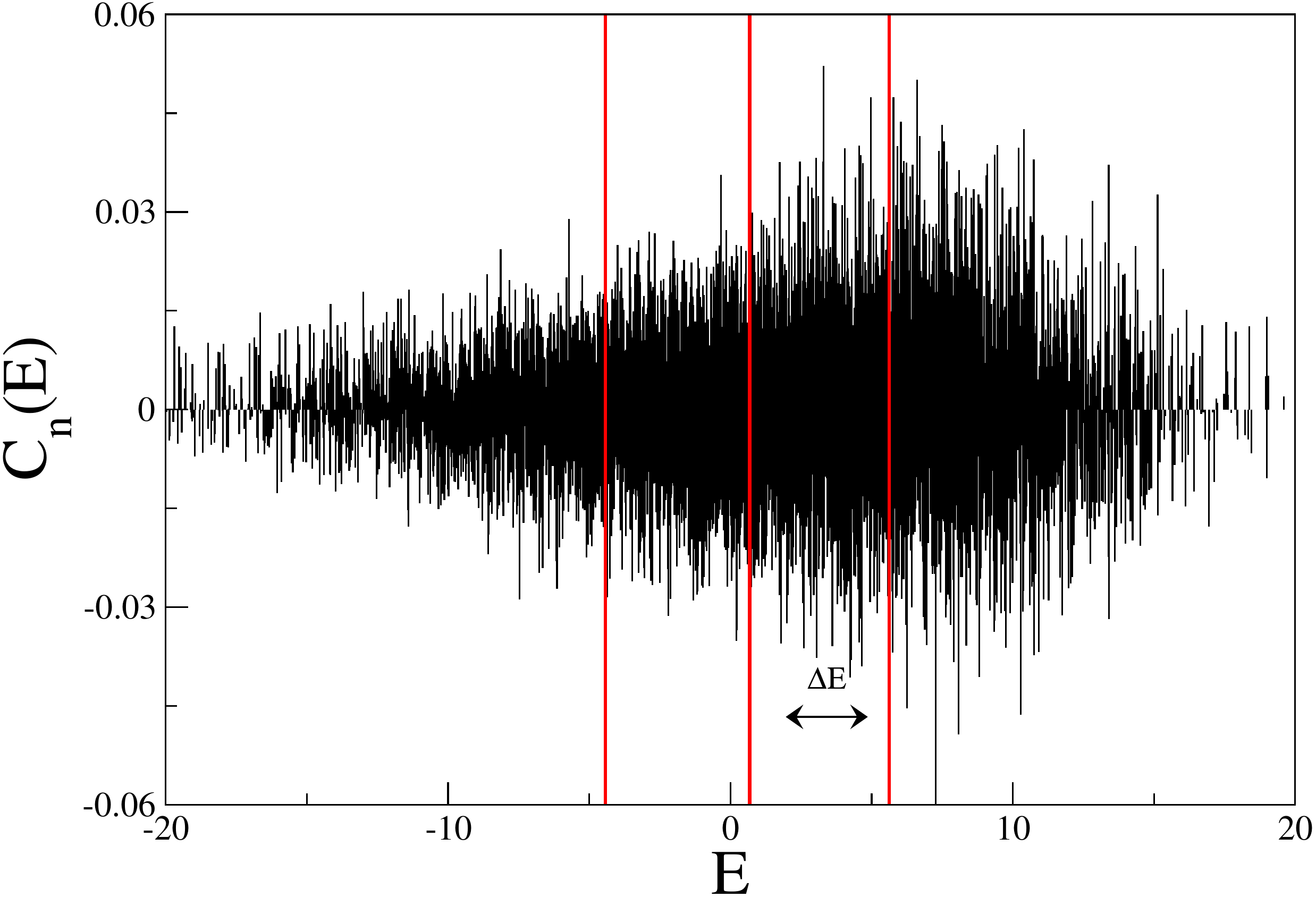}\\
(a)
\end{center}
\end{minipage}
\begin{minipage}{.59\linewidth}
\begin{center}
\includegraphics[width=.95\linewidth, clip]{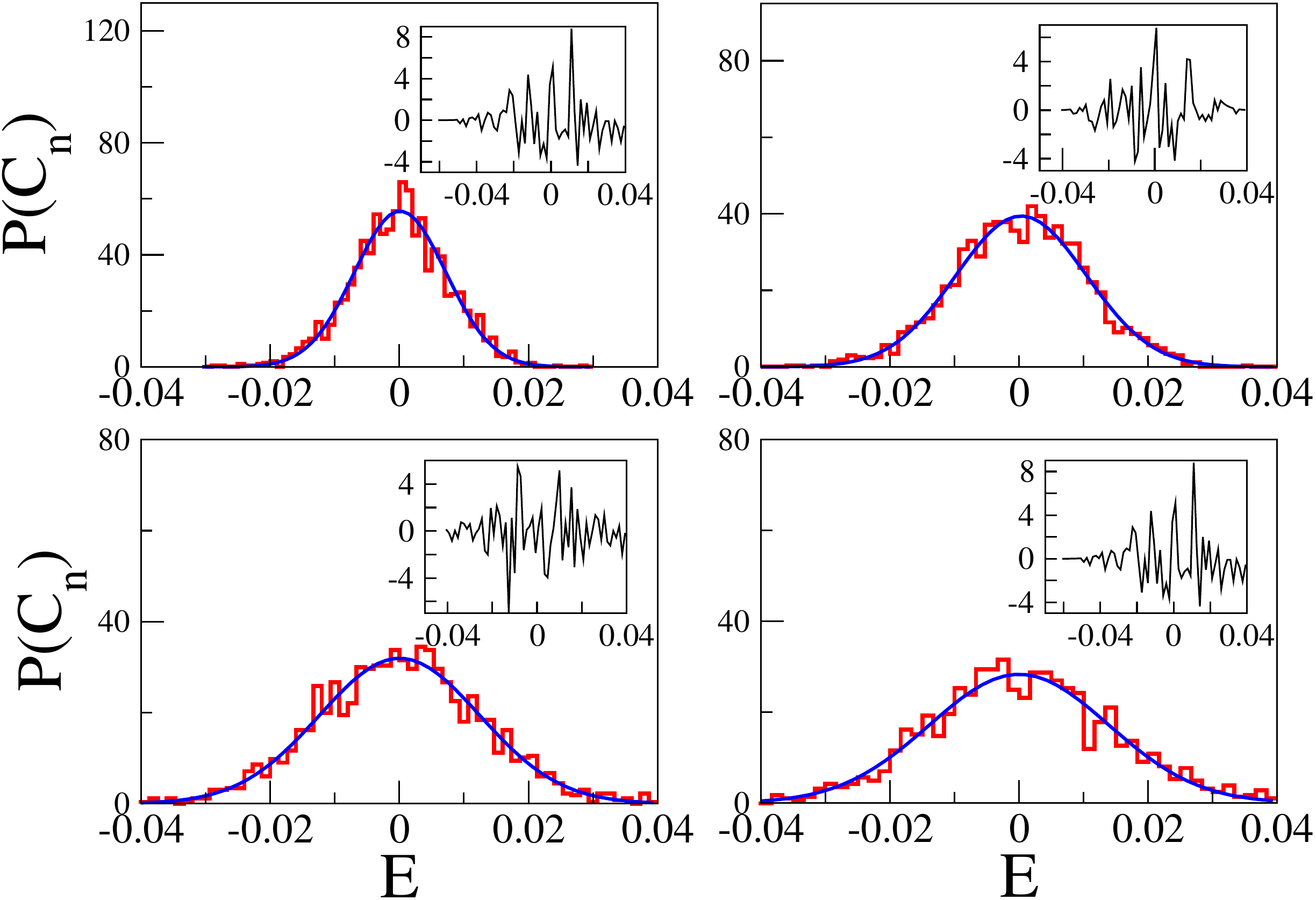}\\
(b)
\end{center}
\end{minipage}
\caption{(a) Coefficient corresponding to symbol $|\, 00001000110111001\,\rangle $ for the Ising model with $\alpha=1$ and $\lambda=1$ for $N=17$ spins in the sector with zero translational momentum. Red vertical lines indicate energy windows used in the calculation of distributions of this coefficient in Fig.~\ref{Gaussian_fits_N_17}b). Each window contains approximately $2000$ levels. 
(b) Distribution of the coefficient as in Fig.~\ref{Gaussian_fits_N_17}a) in 4 energy windows indicated in that figure (red histograms). Blue solid lines are the best Gaussian fits to these distributions. Inserts: The difference between numerical histograms and the Gaussian fits.  }
\label{Gaussian_fits_N_17}
\end{figure}

As an example, we present in Fig.~\ref{Gaussian_fits_N_17}a)  one particular coefficient for all eigen-energies of the Ising model with $\alpha=1$ and $\lambda=1$ for $17$ spins with periodic boundary conditions obtained by direct numerical diagonalization. This type of pictures strongly suggests the following conjecture.    

\textbf{Conjecture 1:} The irregular behaviour of coefficients $C_{\vec{n}}(\alpha)$ in a small energy window can be mimic by 
the assumption that they are random functions of energies with a certain distribution. 

For strongly chaotic systems there are two widely used a-priori assumptions. If coefficients are real,  they should well be described by   the Gaussian distribution. It means that at fixed symbol $\vec{n}$, the distribution density of coefficients is
\begin{equation}
P(C_{\vec{n}}=x)=\frac{1}{\sqrt{2\pi \Sigma_{\vec{n}}^2}}\exp \left (-\frac{x^2}{2\Sigma_{\vec{n}}^2 } \right ).
\label{conjecture_1}
\end{equation}
If coefficients are complex, one assumes that they have Gaussian distribution for the both real and imaginary parts with zero means and the same variance 
\begin{equation}
P(\mathrm{Re}\,C_{\vec{n}}=x,\mathrm{Im}\,C_{\vec{n}}=y)=\frac{1}{\pi \Sigma_{\vec{n}}^2}
\exp \left (-\frac{x^2+y^2}{\Sigma_{\vec{n}}^2 } \right ).
\end{equation}
In the both expressions  $\Sigma_{\vec{n}}^2$ indicates the mean value of modulus square of coefficient $C_{\vec{n}}$
\begin{equation}
\Sigma_{\vec{n}}^2=\langle |C_{\vec{n}}|^2 \rangle 
\end{equation}
For illustration, in Fig.~\ref{Gaussian_fits_N_17}b) we compare distribution of the same  coefficient as in Fig.~\ref{Gaussian_fits_N_17}a) with the best Gaussian fit and a good agreement is clearly seen.

Accepting the above conjecture means, in particular, that  moments of variables $C_{\vec{n}}$ averaged over a small energy interval  $I$ should be  well approximated (at least for large $N$) by the moments of the corresponding Gaussian distributions 
\begin{equation}
\langle |C_{\vec{n}}|^{2q} \rangle_I \equiv \dfrac{\sum_{E_{\alpha}\in I} |C_{\vec{n}}(\alpha)|^{2q} }{\sum_{E_{\alpha}\in I}1}=\mathcal{R}_q\, \Sigma_n^{2q} 
\end{equation}
where for complex coefficients $\mathcal{R}_q=\mathcal{R}_q^{(\mathrm{complex})}$ and for real ones 
$ \mathcal{R}_q=\mathcal{R}_q^{(\mathrm{real})}$ with 
\begin{equation}
\mathcal{R}_q^{(\mathrm{complex})}=\Gamma(q+1), \qquad \mathcal{R}_q^{(\mathrm{real})}= \dfrac{2^q \Gamma\left (q+\tfrac{1}{2}\right ) }{\sqrt{\pi}}.
\label{gaussian_moments}
\end{equation}
 In particular, $\mathcal{R}_2^{(\mathrm{complex})}=2$ and $\mathcal{R}_2^{(\mathrm{real})} =3$.

Under the validity of the conjecture all average moments of $C_{\vec{n}}(\alpha)$ (with fixed symbol $\vec{n}$) are determined by one quantity, the variance $\Sigma_{\vec{n}}^2=\Sigma_{\vec{n}}^2(E)$ which depends on symbol $\vec{n}$ and of the center of energy window $E$.

To calculate this variance, it is convenient to consider the so-called strength function (or the local density of states)
\begin{equation}
P_{\vec{n}}(E)=\sum_{\alpha}|C_{\vec{n}}(\alpha)|^2\delta(E-E_{\alpha}).
\label{strenth_function}
\end{equation}
Due to pseudo-random character of $C_{\vec{n}}(\alpha)$ and $E_{\alpha}$ the strength function itself can be considered as pseudo-random or random function of energy $E$. The mean value of a certain function $f(E)$ is defined as in \eqref{microcanonical}  
\begin{equation}
\Big \langle P_{\vec{n}}(E)f(E)\Big \rangle =\frac{1}{\mathcal{N}_{E,\Delta E}}\sum_{|E-E_{\alpha}|<\Delta E}|C_{\vec{n}}(\alpha)|^2f(E_{\alpha}) 
\label{mean_strength}
\end{equation}
where the width of the energy  window, $\Delta E$,  obeys \eqref{delta_E}.
 
The knowledge of the strength function permits to find $|C_{\vec{n}}(\alpha)|^2$ averaged over a small energy window (which can be used as the definition of the variance)
\begin{equation}
\Sigma_{\vec{n}}^2 =\dfrac{\langle P_{\vec{n}}(E) \rangle }{\rho(E)}
\label{variance}
\end{equation}
where $\rho(E)$ is the average density of states with energy $E$ i.e. the number of states in an interval $E<E_{\alpha}<E+\Delta E$.

The advantage of the strength function is that for this quantity there exist exact sum rules  \cite{b_m}
\begin{equation}
\int P_{\vec{n}}(E)E^k\mathrm{d}E\equiv \langle E^k\rangle =\langle \vec{n}\,  |\,\mathcal{ H}^k\, |\vec{n}\,  \rangle
\label{restrictions}
\end{equation}
It means that these moments can be calculated directly from the Hamiltonian without solving the full problem. 

Of course, the exact calculation of the strength function is equivalent to the full solution of the problem. To obtain a simple approximation one has to assume that the strength function can be well approximated from the knowledge of its first moments. One can argue that for $N$-body systems with short-range interactions the first two moments reproduce well higher order moments when $N\to\infty$.  It leads to the second conjecture:
  
\textbf{Conjecture 2:} The functional dependence  of $P_{\vec{n}}(E)$ on $E$ is well approximated when $N\to\infty$ by a  Gaussian 
\begin{equation}
P_{\vec{n}}(E)=\frac{1}{\sqrt{2\pi \sigma_{\vec{n}}^2}}\exp \left (-\frac{(E-E_{\vec{n}})^2}{2\sigma_{\vec{n}}^2 } \right )
\label{Gaussian}
\end{equation}
where  $E_{\vec{n}}$ and $\sigma_{\vec{n}}^2$ are the first two moments of the Hamiltonian 
\begin{equation}
E_{\vec{n}}=\langle \vec{n}\, |\mathcal{H}|\vec{n}\,  \rangle,\qquad \sigma_{\vec{n}}^2=\langle \vec{n}\,  |(\mathcal{H}-E_n)^2|\vec{n} \, \rangle
\end{equation}
Under this conjecture the variance of  coefficient distribution given by Eq.~\eqref{variance} is
\begin{equation}
\Sigma_{\vec{n}}^2  \approx  \dfrac{1}{\rho(E) \, \sqrt{2\pi \sigma_{\vec{n}}^2}}\exp \left (-\frac{(E-E_{\vec{n}})^2}{2\sigma_{\vec{n}}^2 } \right )
\end{equation}
with the mean spectral density calculated by the formula (to ensure the normalization $\sum_{\vec{n}}|C_{\vec{n}}(\alpha)|^2=1$)
\begin{equation}
\rho(E)=\sum_{\vec{n}}\frac{1}{\sqrt{2\pi \sigma_{\vec{n}}^2}}\exp \left (-\frac{(E-E_{\vec{n}})^2}{2\sigma_{\vec{n}}^2 } \right )
\label{multi_gaussian_density}
\end{equation}
Assuming that coefficients have Gaussian distribution \eqref{conjecture_1}, the values of $2q^{\mathrm{th}}$ moments of the wave function are
\begin{equation}
M_q\equiv \left \langle \sum_{\vec{n}}|C_{\vec{n}}(\alpha)|^{2q}\right \rangle = \frac{\mathcal{R}_q}{\rho^q(E)} \sum_{\vec{n}}
\frac{1}{\big (2\pi \sigma_{\vec{n}}^2\big )^{q/2}}\exp \left (-q\frac{(E-E_{\vec{n}})^2}{2\sigma_{\vec{n}}^2 } \right )
\end{equation}
The above conjectures are basis ingredients of construction of statistical models for chaotic wave functions in different problems \cite{rigol}, \cite{french_1}-\cite{elon}. Though they were not proved in full generality (but see \cite{keating_1}, \cite{keating_2}), they are simple enough to be check for a given particular system. One can even reverse the arguments and say that wave function of $N$-body  model with short range interactions is called  chaotic iff it obeys the above conjectures. 

Another line of reasoning may be the maximum-entropy principle (see e.g. \cite{jaynes}) according to which "the best" statistical distribution ($P_{\vec{n}}(E)$ in our case) fulfilled certain restrictions (as in \eqref{restrictions}) is given by the one which maximise  the entropy 
\begin{equation}
S=-\int P_{\vec{n}}(E)\ln \, P_{\vec{n}}(E)\, \mathrm{d}E
\end{equation}
Assuming  e.g. that a finite number of first moments in \eqref{restrictions} are specified,  the maximum-entropy principle predicts  that the corresponding probability density takes the Gibbsian form 
\begin{equation}
P_{\vec{n}}(E)=\dfrac{\mathrm{e}^{-\sum_j\mu_j E^j}}{Z(\vec{\mu}\,)},\qquad 
Z(\vec{\mu}\,) =\int \mathrm{e}^{-\sum_j\mu_j E^j}\mathrm{d}E  
\label{gibbs_distribution}
\end{equation}
with Lagrangian multipliers, $\mu_j=\mu_j(\vec{n}\,)$ calculated from the partition function using the knowledge of the first moments
\begin{equation}
\langle \vec{n}| \mathcal{H}^j|\vec{n}\rangle =-\frac{\partial}{\partial \mu_j}\ln Z(\vec{\mu}\,)
\end{equation}    
Of course, when only two moments are taken into account one gets \eqref{Gaussian}.  

Though in all considered cases expression \eqref{gibbs_distribution} gives the best results, it requires numerical calculations of  Lagrangian multipliers which complicates the separation of different sources of corrections. To increase the accuracy of approximation, we shall include  the third and the forth  moments  assuming that  that they are much smaller than the first two moments  by using Gram-Charlier A series in Hermite polynomials. It leads to the following modification of $P_n(E)$
\begin{equation}
P_{n}(E)=\frac{1}{\sqrt{2\pi \sigma^2}}\exp \left (-\frac{(E-E_n)^2}{2 \sigma^2} \right ) \left [1+\frac{k_3}{3!\sigma^3} H_3\Big (\frac{E-E_n}{\sigma} \Big )+\frac{k_4}{4!\sigma^4} H_4\Big(\frac{E-E_n}{\sigma} \Big ) \right ]
\label{LDOS_corrected}
\end{equation}
Here $k_3$ and $k_4$ are the third and the forth cumulants for the Ising model given by Eqs.~\eqref{third} and \eqref{forth} from Appendix~\ref{higher_moments},     $H_3(x)=x^3-3x$ and $H_4(x)=x^4-6x^2+3$.  

\section{Participation ratio for the Ising model in two fields}\label{participation} 

For the Hamiltonian \eqref{ising_2} simple calculations give
\begin{equation}
E_{n}=\lambda(N-2n),\qquad \sigma_{n}^2\equiv \sigma^2= N(1+\alpha^2)
\end{equation}
where $n$ is the number of spins up. Each state with fixed $n$ is $C_N^n$ degenerated.

These formulae mean that
\begin{equation}
P_{n}(E)=\frac{1}{\sqrt{2\pi \sigma^2}}\exp \left (-\frac{(E-E_n)^2}{2 \sigma^2} \right )
\label{LDOS_Gaussian}
\end{equation}
Notice that we wrote $n$ and not $\vec{n}$ to stress that in this approximation all quantities depend  only on number of spins up.

\begin{figure}
\begin{center}
\includegraphics[width=.4\linewidth, clip]{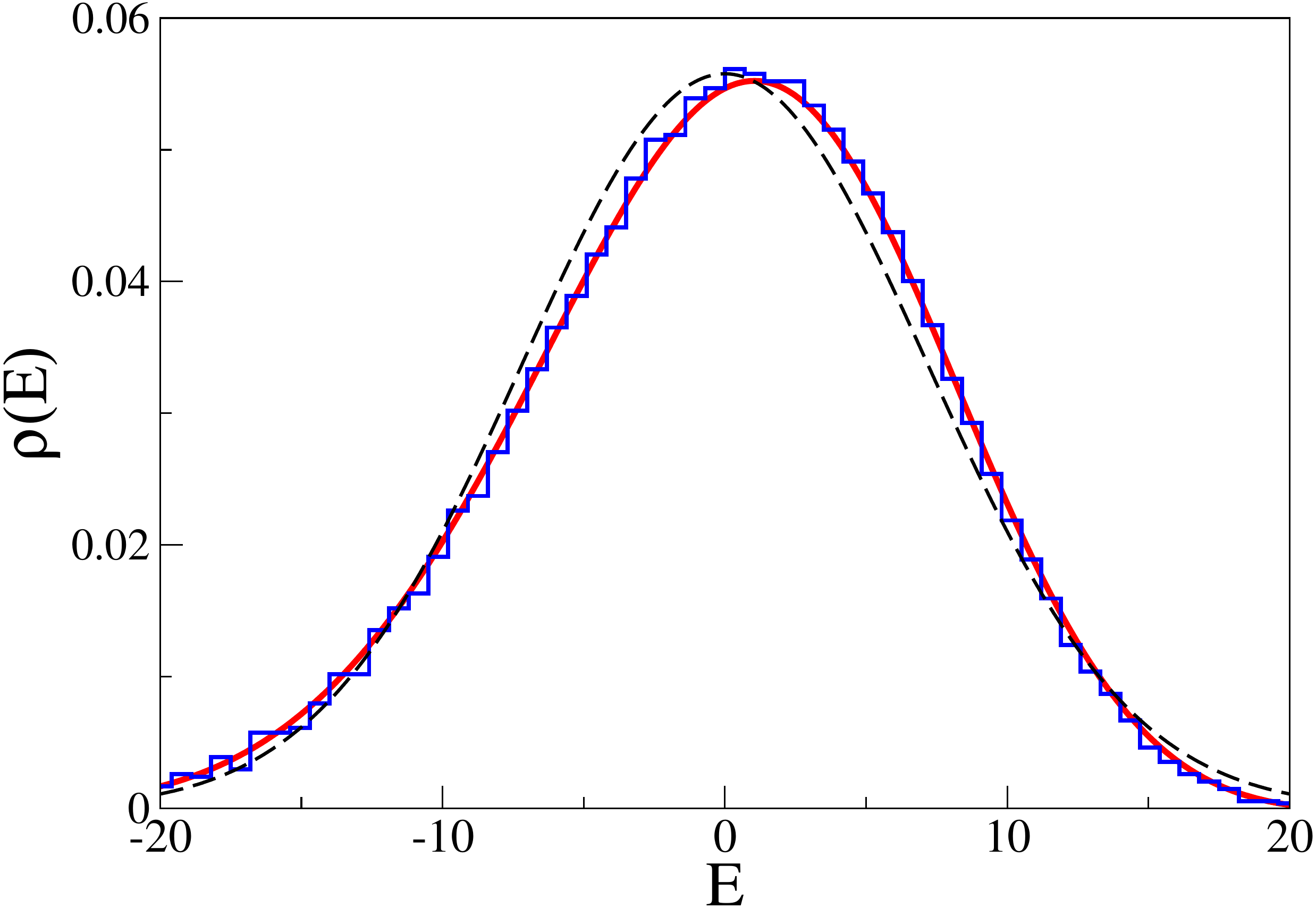}
\end{center}
\caption{Spectral density for the Ising model $\alpha=1$ and $\lambda=1$ for $N=17$ spins. Blue histogram indicates the numerical results. Black dashed line is the Gaussian approximation without corrections \eqref{density_Gaussian} with \eqref{LDOS_Gaussian}.  Red solid line the corrected  approximation \eqref{density_Gaussian} with \eqref{LDOS_corrected}.}
\label{density_Ising_N_17}
\end{figure}

The full density normalized to 1 is
\begin{equation}
\rho(E)=2^{-N} \sum_{n=0}^N C_N^n P_n(E) 
\label{density_Gaussian}
\end{equation}
In Fig.~\eqref{density_Ising_N_17} this equation is compared with the spectral density of the Ising model \eqref{ising_2} with $\alpha=1$ and $\lambda=1$ for $N=17$ spins calculated numerically. The agreement is good but small differences are visible. When higher moments taken into account as in \eqref{LDOS_corrected} the result is practically indistinguishable from numerics.  

To approximate moments of wave functions accurately one has to know not only the first moments of wave function coefficients but also are they real or complex. Usually, the answer is simple. If the Hamiltonian is real symmetric, wave functions (and their coefficients) are real. If the Hamiltonian is complex Hermitian, coefficients  are also complex. For the considered Ising model in two fields \eqref{ising_2} the situation is more tricky. The point is that this model with periodic boundary conditions has the translational invariance
\begin{equation}
\hat{T}:\; n\to n+1,\qquad n=1,2,\ldots,N
\end{equation}
so if $\Psi$ is a wave function with energy $E$,  
$ \Psi=\sum_{n_j=\pm 1}C_{n_1,n_2,\ldots,n_N}|n_1,n_2,\ldots,n_N\rangle$ 
then
$\hat{T}\Psi=\sum_{n_j=\pm 1}C_{n_1,n_2,\ldots,n_N}|n_N,n_1,\ldots,n_{N-1}\rangle$ 
is also an eigenfunction with the same energy which imposes certain relation on coefficients.

As $\hat{T}^N=1$, its eigenvalues are $\mathrm{e}^{2\pi\mathrm{i}k/N}$  and  eigenfunctions can be classified by the translational  momentum $k=0,1,\ldots, N$ 
\begin{equation}
\hat{T}\Psi_k=\mathrm{e}^{2\pi\mathrm{i}k/N}\Psi_k
\label{translational_states}
\end{equation}
Numerical calculations below are performed in the basis with fixed translational momentum $k$. 

To construct explicitly such basis one fixes one element, $|\vec{n}\, \rangle$, $\vec{n}=(n_1,n_2,\ldots,n_N)$ and constructs all its translations with corresponding phases. For prime $N$ one can choose
\begin{equation}
|\vec{n}\, \rangle_k=\frac{1}{\sqrt{N}}\sum_{j=0}^{N-1}\mathrm{e}^{2\pi \mathrm{i}kj/N}\hat{T}^j|\vec{n}\, \rangle
\label{sum_k}
\end{equation}
Here $\hat{T}^j|\vec{n}\, \rangle$ indicates  the shift on $j$ elements of the initial sequence
\begin{equation}
\hat{T}^j|n_1,n_2,\ldots,n_N\rangle\equiv |n_{N-j+1},\ldots,n_N,n_1,n_2,\ldots, n_{N-j}\rangle 
\end{equation}
After performing one step  one takes arbitrary  element which is not in the sum \eqref{sum_k} and repeats the construction till all elements are exhausted.  

For composite $N$ there exist elements with primitive periods $t$ equals to a divisor of $N$. Their number is
\begin{equation}
r(t)=\frac{1}{t}\sum_{d|t}2^{t/d}\mu(d)
\end{equation} 
where $\mu(n)$ is the M\"obius function, $\mu(n)=(-1)^{n_p}$ if $n$ is square-free number with $n_p$ prime divisors, and $\mu(n)=0$ if $n$ is divisible on a square of a prime. 

The total  dimension of the basis with $k=0$ is
\begin{equation}
\mathcal{N}_{\mathrm{tot}}=\sum_{n|N}r(n)=\frac{1}{N}\sum_{d|N}2^{N/n}\phi(d)
\end{equation}
where $\phi(n)$ is Euler's totient function equals the number of positive integers smaller than $n$ and co-prime with it.           
For non-zero $k$ the contribution of basis states with primitive period $t<N$, ($t|N$), is non-zero only when $k\equiv 0$  mod $N/t$. 

The simplest approximation 
\begin{equation}
\mathcal{N}_{\mathrm{tot}}\approx \dfrac{2^N}{N}
\label{total_N}
\end{equation}
is sufficient  in many cases when $N$ is large. The total number of states with fixed momentum and fixed number of spins up $\nu_{\mathrm{tot}}(n)$ can also be approximated in the similar manner
\begin{equation}
\nu_{\mathrm{tot}}(n)\approx \frac{1}{N}C_N^n
\label{nu_tot}
\end{equation} 
After rewriting the Hamiltonian in the basis of states with fixed translational momentum, $|\vec{n}_k\rangle$, it becomes complex and eigenfunctions
\begin{equation}
\Psi_k=\sum_{\vec{n}_k} C_{\vec{n}_k}|\vec{n}\rangle_k 
\label{psi_k}
\end{equation}
are, in general, also complex.

Nevertheless, one cannot conclude that in the chaotic regime eigenstates and eigenfunctions are distributed as for Gaussian Unitary ensemble of random matrices (GUE).  The reason is that  for the model considered there exists another discrete symmetry, namely geometric inversion
\begin{equation}
\hat{S}:\, n\to N-n+1,\qquad n=1,2,\ldots, N
\end{equation}
Hamiltonian \eqref{ising_2} is invariant under this transformation. Therefore if $\Psi$ is an eigenfunction $\hat{S}\Psi$ is also an eigenfunction with the same energy.

By definition the initial basis $|\vec{n}\,\rangle$  transforms under $\hat{S}$ as follows
\begin{equation}
\hat{S}|n_1,n_2,\ldots,n_N\rangle = |n_N,n_{N-1},\ldots,n_1\rangle
\end{equation}
Functions with zero translational momentum transform into themselves under this inversion 
\begin{equation}
\hat{S}\Psi_{0}=\epsilon \Psi_{0}
\end{equation} 
where $\epsilon=\pm 1$ is the parity under the inversion. But functions with $k\neq 0$ are transformed up to a phase into functions with opposite momentum
\begin{equation}
\hat{S} \Psi_{k}= \Psi_{N-k}
\end{equation} 
States $|\vec{n}\rangle_k$ with fixed momentum $k$ (cf. Eq.~\eqref{sum_k}) split into two groups under the inversion. The first group includes all states which under this inversion transform into  themselves but with opposite momentum (once more up to a certain non-dynamical phase which can be included in the definition of basis states)
\begin{equation}
\hat{S}|\vec{n}\, \rangle_k=|\vec{n}\, \rangle_{N-k}
\end{equation}
In other words, the inversion of one element of this group is equivalent to a certain shift of this element. Such elements are called invariant under the inversion. 

Elements of the second group are organised in different pairs transformed one to another by the inversion
\begin{equation}
\hat{S}|\vec{n}\, \rangle_k= |\vec{n}^{\prime}\, \rangle_{N-k},\qquad \vec{n}^{\prime}\neq n
\end{equation}
We call them non-invariant elements. 

The above arguments show that eigenfunctions with $k\neq 0,N/2$  can be written in the form
\begin{equation}
\Psi_k=\sum_{\mathrm{non-invariant}}\left (C_{\vec{n}_k}^{(\mathrm{n-in})}|\vec{n}\, \rangle_k+C_{\vec{n}_k}^{(\mathrm{n-in})\, *} |\vec{n}^{\prime}\, \rangle_k\right )+
\sum_{\mathrm{invariant}} C_{\vec{n}_k}^{(\mathrm{in})}|\vec{n}\, \rangle_k
\label{sector_non_zero}
\end{equation}
where $\vec{n}$ and $\vec{n}^{\prime}$ are pairs of elements connected by the inversion. Taking real and imaginary parts of this expression demonstrates that all coefficients in the expansion can be chosen real which explains that spectral statistics of this model is well described by GOE statistics (either give the reference or a picture of left-right asymmetry for non-zero momentum!) though the Hamiltonian itself is complex (cf. \cite{triangle}). 

For $k=0$ eigenfunctions  have a particular parity under the inversion  and all coefficients can be chosen real but invariant elements are identically zero for states with negative parity
\begin{equation}
\Psi_0^{(\epsilon)}=\sum_{\mathrm{non-invariant}}\left (C_{\vec{n}}^{(\mathrm{n-in})}|\vec{n}\, \rangle_0+\epsilon C_{\vec{n}}^{(\mathrm{n-in})} |\vec{n}^{\prime}\, \rangle_0\right )+
\tfrac{1+\epsilon}{2}\sum_{\mathrm{invariant}} C_{\vec{n}}^{(\mathrm{in})}|\vec{n}\, \rangle_0
\label{sector_zero}
\end{equation}
with $\epsilon=\pm 1$. 

It is plain that the majority of elements at large $N$ are non-invariant. Nevertheless,  corrections from invariant elements, though small,  are noticeable for $N$ accessible in numerical calculations.  The number of invariant elements is especially simple to find for prime $N$ when all elements have the same primitive period.  As $N=2K+1$ is odd, all invariant elements in the periodic cycle are appeared in pairs under the inversion except one element which has to be invariant under the inversion. Such element can  be constructed by choosing arbitrary the first $K$ elements and reflecting them into another part of the element. The number of possibilities is $2^K$ and one has $2$ chooses for the central element, therefore the total number of invariant elements is 
\begin{equation}
\mathcal{N}_{\mathrm{inv}}=2^{K+1}=2^{[N/2]+1}
\label{inv_N}
\end{equation}
where $[x]$ is the integer part of $x$. The number of invariant elements with fixed total number of spins  is
\begin{equation}
\nu_{\mathrm{inv}}(n)=C_{[N/2]}^{[n/2]}
\end{equation}
Similar expression may be obtained for even $N$.

According to the above statistical conjectures, coefficients $C_{\vec{n}_k}^{(\mathrm{n-in})}$ with $k\neq 0$ are distributed as complex Gaussian variables with zero mean and  variance $\Sigma_n^2$ calculated from $P_n(E)$ as in Eq.~\eqref{variance}
\begin{equation}
\Sigma_{\vec{n}}^2 =\dfrac{P_{\vec{n}}(E)}{\rho(E)},\qquad \rho(E)=\sum_{n=0}^N \nu_{\mathrm{tot}}(n) P_n(E) 
\end{equation}
with $P_n(E)$ as in Eqs.~\eqref{LDOS_Gaussian} or \eqref{LDOS_corrected}. 

Similarly, coefficients of invariant terms, $C_{\vec{n}_k}^{(\mathrm{in})}$, have real Gaussian distribution with the same variance.  Therefore, when moments are calculated, one has to take into account corrections from different types of distributions (cf. Eqs.~\eqref{gaussian_moments}). For large $N$ one can neglect the existence of invariant elements and consider that for non-zero $k$ functions are complex and for zero momentum functions are real.

As a typical example we consider the participation ratio determined as usual
\begin{equation}
Pr(E_{\alpha})=\left ( \sum_{\vec{n}} |C_{\vec{n}}(E_{\alpha})|^4 \right )^{-1}
\end{equation}
The simple Gaussian approximation of this quantity takes the form
\begin{equation}
Pr(E)\approx \mathcal{N}_{\mathrm{tot}}\dfrac{\left (2^{-N}\sum_{n=0}^N C_N^n P_n (E) \right )^2 }{\mathcal{R}_2 \left ( 2^{-N}\sum_{n=0}^N C_N^n P_n (E)^2 \right )}
\label{pr_Gaussian}
\end{equation}
where $P_n(E)$ is given by Eq.~\eqref{LDOS_Gaussian}. Here we use approximation \eqref{nu_tot} for the number of total terms with fixed $n$, $\mathcal{N}_{\mathrm{tot}}=\sum_{n}\nu_{\mathrm{tot}}(n)=2^N/N$, $\mathcal{R}_2=2$ for states with non-zero momentum and $\mathcal{R}_2=3$ for zero momentum state.   
 
\begin{figure}
\begin{minipage}{.49\linewidth}
\begin{center}
\includegraphics[width=.95\linewidth, clip]{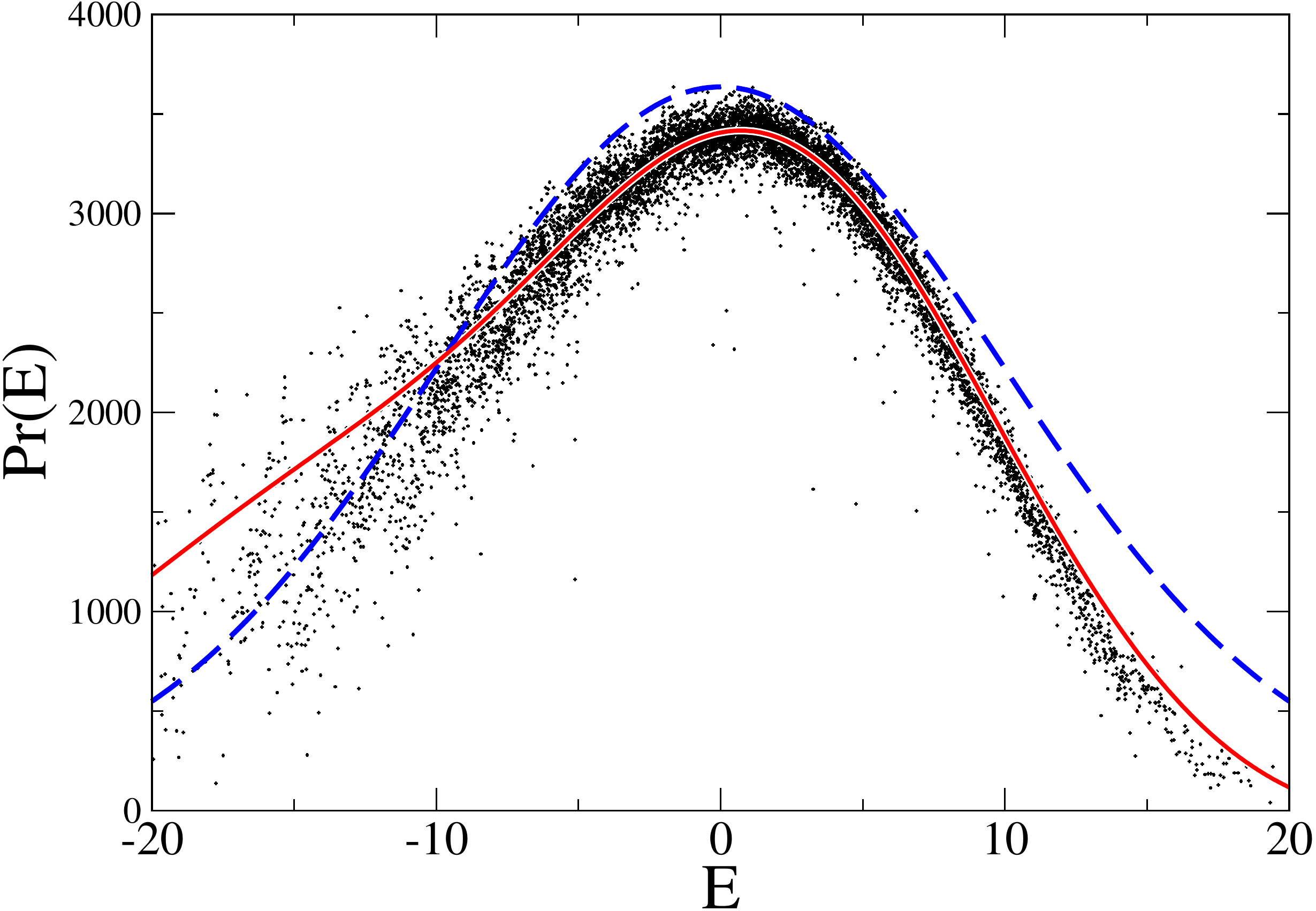}\\
(a)
\end{center}
\end{minipage}
\begin{minipage}{.49\linewidth}
\begin{center}
\includegraphics[width=.85\linewidth, clip]{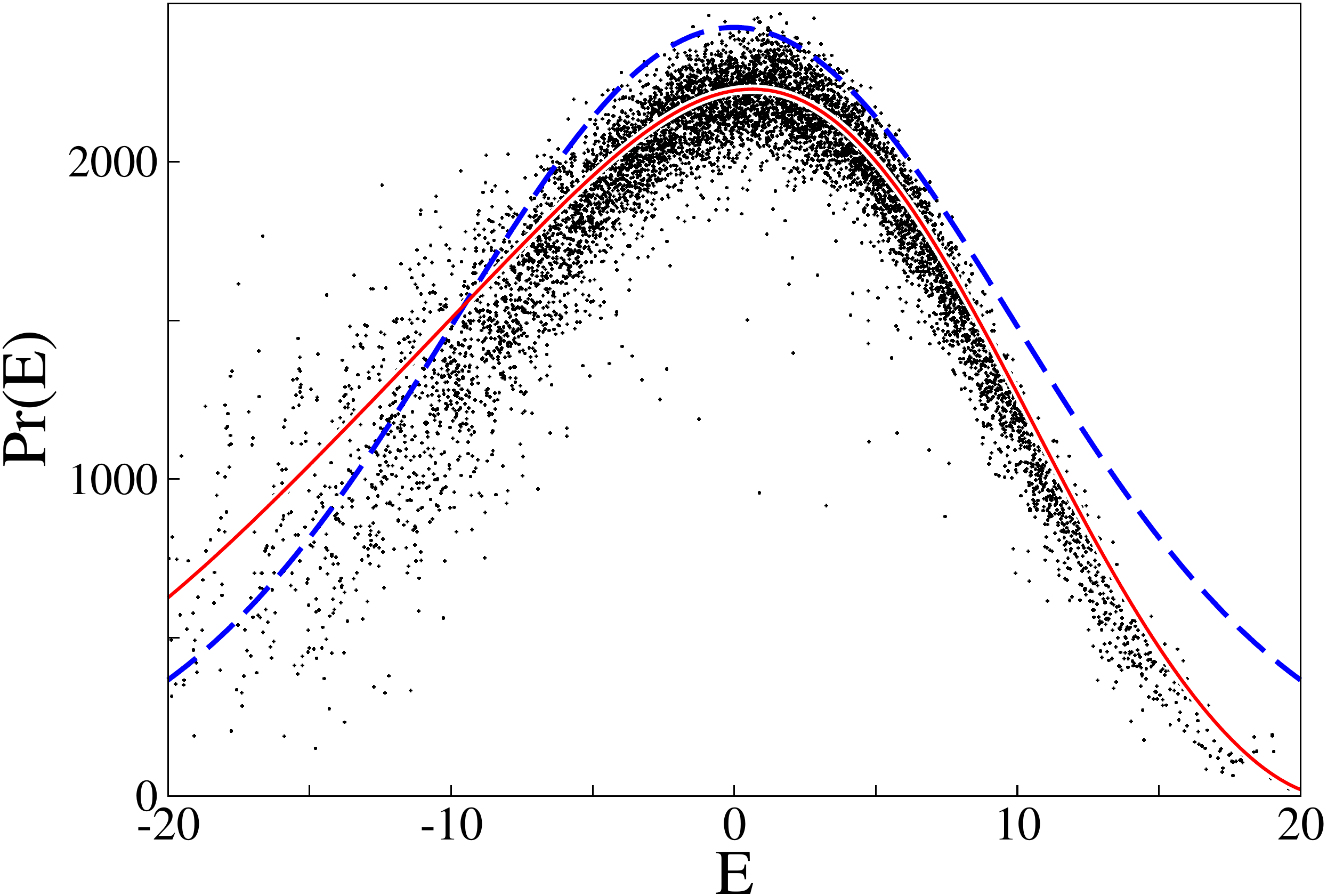}\\
(b)
\end{center}
\end{minipage}
\caption{(a) Participation ratio  of the Ising model with $\alpha=1$ and $\lambda=1$ for $N=17$ spins in sector with translational moment equals 2. Black crosses are result of numerical calculations.   Blue dashed line is the pure Gaussian approximation without corrections \eqref{pr_Gaussian} with $\mathcal{R}_2=2$. Red line is the corrected approximation. 
(b) The same as in (a) but  with zero translational moment. Blue dashed line is the pure Gaussian approximation without corrections \eqref{pr_Gaussian} but with $\mathcal{R}_2=3$. }
\label{participation_ratio_complex_real}
\end{figure}

In Figs.~\ref{participation_ratio_complex_real}a) and \ref{participation_ratio_complex_real}b) we present the numerical results for this quantity for the Ising model with $\alpha=\lambda=1$ and $N=17$ spins for states with translational momentum $k=2$ and $k=0$ respectively.   
Notice that left-hand side of the participation ratio (with $E<-10$) fluctuate strongly and the using statistical description at energies close to the ground state is questionable. In the same figures the above Gaussian approximation is also indicated. In general the agreement is good but there exit noticeable differences with numerical calculations. 

There exist two different types of corrections. The first is related with higher moments of Hamiltonian and they can be taken into account by modifications indicated in Eq.~\eqref{LDOS_corrected}. This formula differs from the Gibbs-like formulae \eqref{gibbs_distribution}. The latter usually give better results, but as they require numerical calculations to find necessary Lagrangian multiplies, their use leads to non-transparent expressions where relative importance of different terms is hidden. To get clear separation of different contributions we prefer to use series expansion as in Eq.~\eqref{LDOS_corrected} though they lead to big errors at large arguments. As in all cases the advocated statistical approach can, strictly speaking, be applied only in the bulk of the spectrum  it is not an important restriction.    Another corrections are related with the fact that in the expansion of eigen functions there exits basis sets which are invariant under inversion (cf. Eqs.{\eqref{sector_non_zero} and \eqref{sector_zero})which has properties different from non-invariant terms.    

For the moments of wave function coefficients with non-zero momentum one gets the following formula
\begin{eqnarray}
\left \langle \sum_{\vec{n}}|\Psi_{\vec{n}}(E)|^{2q}\right \rangle =\left (\mathcal{R}_{q}^{\mathrm{complex}} + [\mathcal{R}_{q}^{\mathrm{real}}-\mathcal{R}_{q}^{\mathrm{complex}}]\delta_q(E)\right )
\dfrac{\sum_{n=0}^N \nu_{\mathrm{tot}}(n) P_n^q (E)  }{ \left ( \sum_{n=0}^N\nu_{\mathrm{tot}}(n) P_n (E)\right )^q}
\end{eqnarray}
where
\begin{equation}
\delta_q(E)=\dfrac{\sum_{n}\nu_{\mathrm{inv}}(n)\, P_n^q(E)}{\sum_{n}\nu_{\mathrm{tot}}(n)\, P_n^q(E)}\approx \frac{\mathcal{N}_{\mathrm{inv}}}{\mathcal{N}_{\mathrm{tot}}}
\end{equation}
The approximative expression of $\delta_q\approx \delta =\mathcal{N}_{\mathrm{inv}}/\mathcal{N}_{\mathrm{tot}}$ corresponds to the uniform approximation where the summation over invariant set is proportional to the summation over all elements.  
When the invariant elements are taken into account in Eq.~\eqref{pr_Gaussian} for the participation ratio  one has  instead of  $\mathcal{R}_2=2$  to use  $\mathcal{R}_2=2+ \delta $ for states with non-zero translational momentum. 

For $k=0$ all eigenfunctions are either symmetric or anti-symmetric with respect to inversion. The full dimensions of  symmetric and anti-symmetric sub-spaces (i.e. the number of independent real coefficients) are (cf. \eqref{sector_zero})  
\begin{equation}
N_{\pm}=\frac{1}{2}(1\pm \delta) \mathcal{N}_{\mathrm{tot}}
\end{equation}  
Eigenvalues corresponding to the both sub-spaces are statistically independent, so the local densities of these eigenvalues are proportional to the above numbers, 
\begin{equation}
\rho_{\pm}(E)=\frac{1}{2}( 1\pm \delta  ) \rho(E)
\label{density_parity}
\end{equation}
Correspondingly, the strength function \eqref{strenth_function} has  contributions from the both symmetric and anti-symmetric states.   According to the above conjecture each term in Eq.~\eqref{sector_zero} has  real Gaussian distribution with the same variance. But this variance are depended on the number of independent components. In the uniform approximation  variances in different sub-spaces are
\begin{equation}
\tilde{\Sigma}_{\vec{n}_0}^2(\pm)=\frac{2}{1\pm \delta}\Sigma _{\vec{n}_0}^2
\end{equation}
For positive parity states the variance of $C_{\vec{n}}^{(\mathrm{in})}$ equals $\tilde{\Sigma}_{\vec{n}_0}^2(+)$ but variance of 
non-invariant coefficients, $C_{\vec{n}}^{(\mathrm{n-in})}$ is,  according to our definition \eqref{sector_zero}, equal to 
 $\tfrac{1}{2}\tilde{\Sigma}_{\vec{n}_0}^2(+)$. Calculating the $2q^{\mathrm{th}}$ moments of eigen-functions in the initial spin basis gives
 \begin{equation}
 M_q^{(+)}=\frac{1-\delta+2^q \delta}{(1+\delta)^q}M_q,\qquad M_q^{(-)}=\frac{1}{(1-\delta)^q}M_q
 \end{equation} 
 As we do not separate states with definite parity, the mean value of moments is (cf. \eqref{density_parity}) 
 \begin{eqnarray}
 \left \langle \sum_{\vec{n}}|\Psi_{\vec{n}}(E)|^{2q}\right \rangle &=&\tfrac{1}{2}( 1+ \delta  )M_q^{(+)}+\tfrac{1}{2}( 1- \delta  )M_q^{(-)}\nonumber\\
 &\approx &\mathcal{R}_q^{\mathrm{real}}\left [1+(2^{q-1}-1)\delta \right ]\dfrac{\sum_{n=0}^N \nu_{\mathrm{tot}}(n) P_n^q (E)  }{ \left ( \sum_{n=0}^N\nu_{\mathrm{tot}}(n) P_n (E)\right )^q} 
 \label{P_q_real}
 \end{eqnarray} 
 The last equation is valid in the first order on $\delta$. one can performed exact summation and with the same accuracy the only difference is the substitution  $\delta\to\delta_q$. 
In particular, the participation ratio is given by Eq.~\eqref{pr_Gaussian} but instead of  $\mathcal{R}_2=3$ one has to use 
$\mathcal{R}_2=3(1+\delta)$.

Inverse moments for the Ising model with the same values of parameters as above  calculated numerically for zero momentum sector are plotted in Fig.~\ref{inverse_moments} for different values of $q$ and compared with Eq.~\eqref{P_q_real}.  Participation ratios for different values of $\alpha$, $\lambda$, and $N$ are shown in Fig.~\ref{different_N}. In all cases in the bulk a good agreement with statistical model with corrections is found in the bulk of the spectrum. 

\begin{figure}
\begin{center}
\includegraphics[width=.95\linewidth, clip]{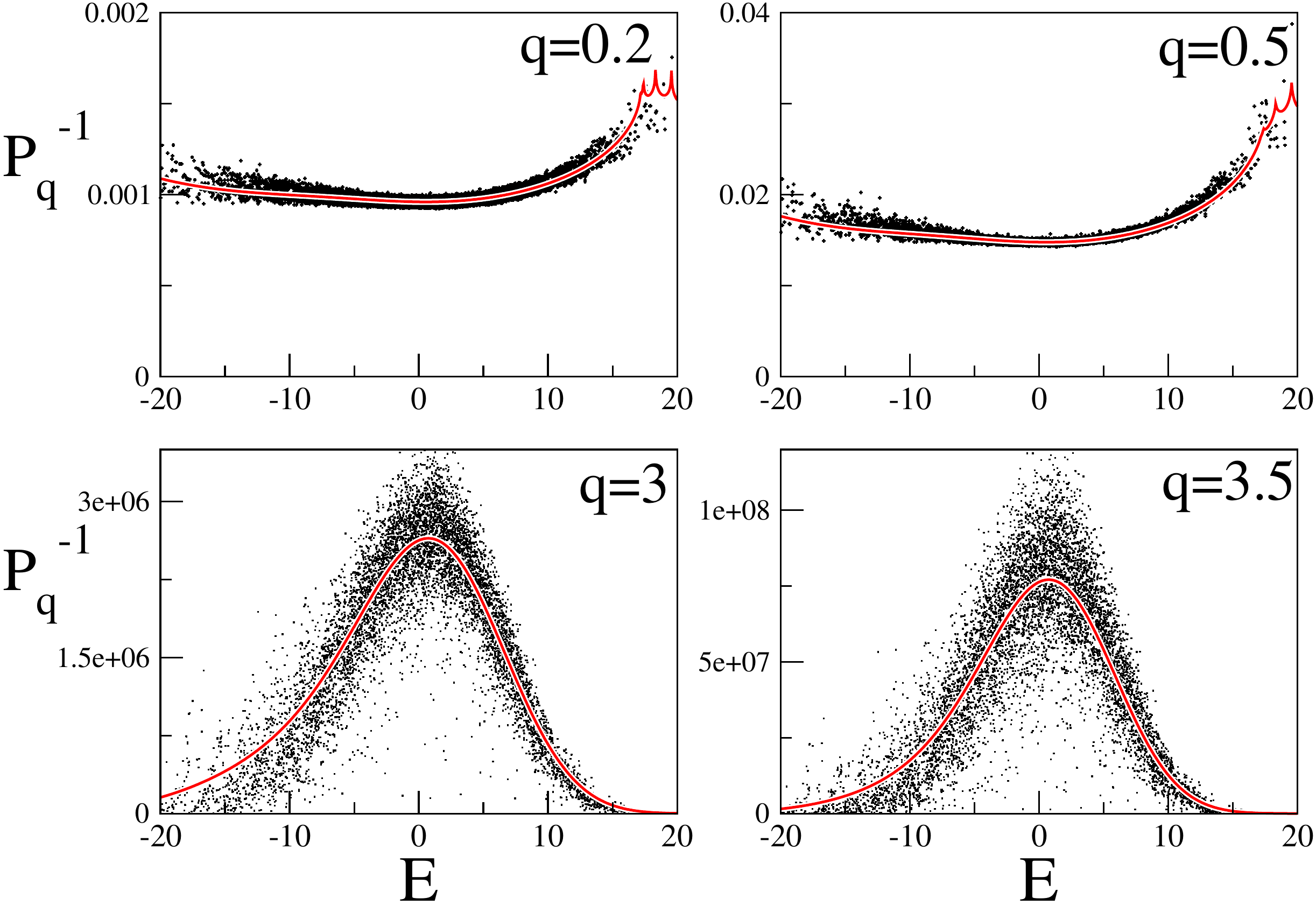}
\end{center}
\caption{Inverse moments of the Ising model with $\alpha=1$ and $\lambda=1$ for $N=17$ spins  with zero translational momentum for different values of $q$. Black crosses are results of numerical calculations.   Red lines are the corrected Gaussian approximation.  }
\label{inverse_moments}
\end{figure}

\begin{figure}
\begin{center}
\includegraphics[width=.95\linewidth, clip]{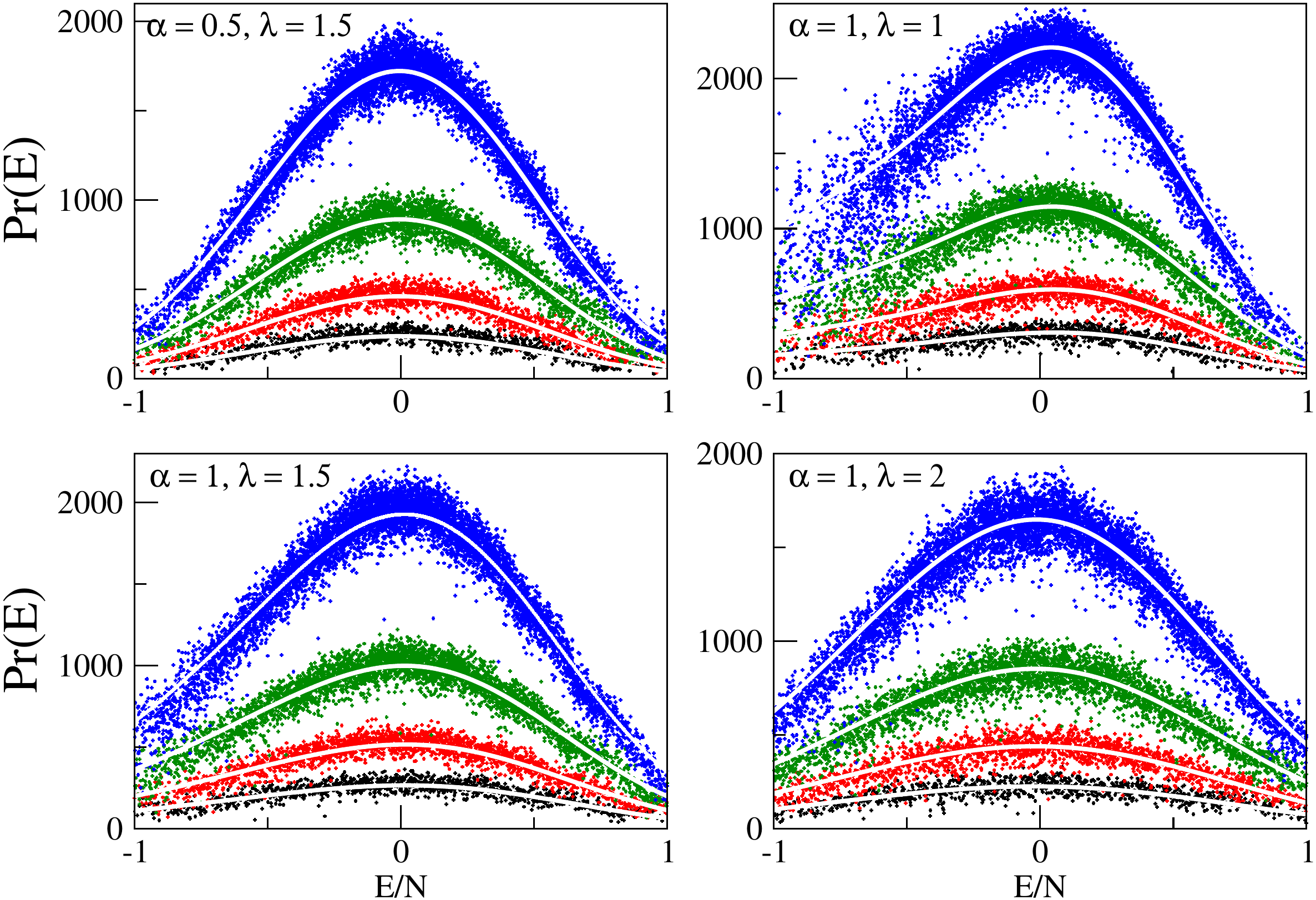}
\end{center}
\caption{Participation ratio for the Ising model for indicated values of  $\alpha$ and $\lambda$ and different $N$ in zero translational momentum sector. From top to bottom $N=17$, $N=16$, $N=15$, and $N=14$. Small crosses are numerical results. White lines are the corrected statistical approximation.  }
\label{different_N}
\end{figure}

\section{Conclusion}

Exact determination of wave functions in many-body problems is rarely possible. For overwhelming majority of models numerical calculations remain the only way of getting information of wave functions. In certain models (called chaotic) the complexity of wave functions is so high that with a good precision they can be considered as random and the developing of statistical methods becomes a  valuable alternative to numerics. This in turn enables meaningful formulations of questions about the foundations of quantum thermodynamics, which is a work in progress in modern physics. 

In the paper we discuss the construction of statistical models for eigenfunctions in a particular example of one-dimensional models of $N$ interacting spin-$\tfrac{1}{2}$, namely  the Ising model in transverse and longitudinal fields. The model is simple enough to demonstrate general phenomena without unnecessary complications. 

The investigation of the model is restricted to the  chaotic regime when all coupling constants are of the same order. It is attested that for large number of spins eigenfunction coefficients in the bulk of the spectrum are well approximated by Gaussian functions with zero mean and variance determined analytically from the Hamiltonian. Such asymptotic results are supposed to give a good description of wave functions only in thermodynamic limit when the number of spins tends to infinity. For numbers of spins accessible in numerical calculations there exist small but noticeable deviations from  asymptotic formulae and a large part of the paper is devoted to  calculations of different types of corrections.   

The first type of corrections is common. It appears every time one calculates corrections to the central limit theorem by taking into account higher order moments.  The best way to incorporate them is to use Gibbs-like formulae which requires numerical calculations to find necessary parameters. When the number of spins is large, contributions of higher order cumulants  are small and one can incorporate  these corrections by using simple series in the Hermite polynomials.      
 
The second type of corrections appears for models with periodic boundary conditions and is related with  the conservation of translational momentum and  the parity transformation.  Due to symmetry considerations some coefficients are real and others have to be complex. In addition, for certain values of translational momentum the symmetries impose that a subset of coefficients is zero.     

When all corrections are included, the resulting statistical expressions agree well with the results of numerical calculations of  
of wave function moments which clearly confirms that in the bulk wave functions are fully extended. 

All constructions used in the paper can be applied only in the bulk of the spectrum where eigen-energies  are of the order of the square root of number of spins. The structure of eigenfunctions close to the ground state and to the highest energy state may and will be more complicated. In particular, the ground state eigenfunctions for practically all spin chains have non-trivial multifractal dimensions \cite{b_a}. The existence or not a sharp transition between multifractal behaviour of boundary states and the extended states in the bulk and more careful determination of  the nature of these states (glassy?)  require additional investigations.  The behaviour of wave functions for small or large values of  coupling constants in the bulk is different from the one investigated so far.  The existence in such cases of perturbation series for large but finite number of spins leads to multiple peaks in the both average spectral density \cite{a_b} and in moments of wave functions. These questions will be discussed elsewhere.    
\appendix
\section{Calculations of first moments for the Ising model}\label{higher_moments}

Hamiltonian of the Ising model in two fields \eqref{ising_2} can be rewritten in the form 
\begin{equation}
\mathcal{H}=\mathcal{H}_x+\mathcal{H}_z
\label{h_x_h_z}
\end{equation}
where
\begin{equation}
\mathcal{H}_x=-\sum_{n}h_n,\qquad h_n= \sigma_n^x\sigma_{n+1}^x+\alpha\,  \sigma_n^x,\qquad 
\mathcal{H}_z= -\lambda\sum_n\sigma_n^z
\end{equation}
The purpose of this Appendix is to calculate  moments of the Hamiltonian 
$\mu_k(\vec{n})\equiv \langle \vec{n}| \mathcal{H}^k|\vec{n}  \rangle$ 
taken between  basis states with definite projection of spins in each point. 
 
As  $\mathcal{H}_z|\vec{n}  \rangle=E_n|\vec{n}  \rangle$ with $E_n=\lambda (N-2n)$
where $n$ is the number of spins up, one has $\langle \vec{n}| \mathcal{H}_z^k| \vec{n} \rangle = E_n^k$. 
Therefore the first two moments of the Ising model Hamiltonian are
\begin{equation}
\mu_1(n)=E_n,\qquad \mu_2(n)=N(1+\alpha^2)+E_n^2
\end{equation}
and the variance (i.e. the second cumulant)
\begin{equation}
\sigma_n^2=\langle \vec{n}| \mathcal{H}^2|\vec{n}  \rangle-\langle \vec{n}| \mathcal{H}|\vec{n}  \rangle^2=N(1+\alpha^2)
\end{equation}
Hamiltonian $\mathcal{H}_x$ flips spins so simple counting gives 
\begin{equation}
\langle \vec{n}| \mathcal{H}_x|\vec{n}  \rangle=0,\qquad \langle \vec{n}| \mathcal{H}_x^2|\vec{n}  \rangle=N(1+\alpha^2),\qquad 
\langle \vec{n}| \mathcal{H}_x^3|\vec{n}  \rangle = -6N\alpha^2
\label{x_moments}
\end{equation}
The forth power of $\mathcal{H}_x$ is
\begin{eqnarray}
\mathcal{H}_x^4&=&\sum_n h_n^4+4\sum_{n_1\neq n_2}h_{n_1}^3 h_{n_2}+6\sum_{n_1<n_2}h_{n_1}^2h_{n_2}^2+12\sum_{n_1\neq n_2<n_3}h_{n_1}^2h_{n_2}h_{n_3}\nonumber\\
&+& 24\sum_{n_1< n_2<n_3< n_4}h_{n_1}h_{n_2}h_{n_3}h_{n_4}
\end{eqnarray}
By inspection one can check that only the even powers of $h_n$ give contribution to  
$\langle \vec{n}| \mathcal{H}_x^4|\vec{n}  \rangle$  and 
\begin{equation}
\langle \vec{n}| \mathcal{H}_x^4|\vec{n}  \rangle =3N^2(1+\alpha^2)^2+N(24\alpha^2-2-2\alpha^4)
\end{equation}
As it does not depend on $n$ it coincides with $1/2^N$Tr~$\mathcal{H}_x^4$ \cite{a_b}.

The third power of the Hamiltonian is 
\begin{equation}
\mathcal{H}^3 =\mathcal{H}_x^3+\mathcal{H}_x^2 \mathcal{H}_z+\mathcal{H}_x \mathcal{H}_z\mathcal{H}_x+\mathcal{H}_x\mathcal{H}_z^2
+\mathcal{H}_z\mathcal{H}_x^2+\mathcal{H}_z\mathcal{H}_x\mathcal{H}_z+\mathcal{H}_z^2\mathcal{H}_x+\mathcal{H}_z^3
\end{equation}
Due to Eq.~\eqref{x_moments} one can find all contributions to $\langle \vec{n}| \mathcal{H}^3|\vec{n}  \rangle$  except $\mathcal{H}_x \mathcal{H}_z\mathcal{H}_x$. 

One gets
\begin{equation}
\langle \vec{n}| \mathcal{H}^3 |\vec{n}  \rangle=
\langle \vec{n}| \mathcal{H}_x^3|\vec{n}  \rangle +
2E_n \langle \vec{n}| \mathcal{H}_x^2|\vec{n}  \rangle + 
3E_n^2 \langle \vec{n}| \mathcal{H}_x |\vec{n}  \rangle + E_n^3 +
\langle \vec{n}|\mathcal{H}_x \mathcal{H}_z\mathcal{H}_x|\vec{n}\rangle
\end{equation}
As has been discussed above, $\mathcal{H}_x$ acting on a product state  $|\vec{n}\rangle$ changes the number of spins up which can be symbolically written as follows 
\begin{eqnarray}
\mathcal{H}_x&\underset{n\to n}{\longrightarrow}& - 1, \;2k\ \mathrm{times} \label{n_minus}\\
\mathcal{H}_x&\underset{n\to n +1}{\longrightarrow}&  -\alpha, \; (N-n)\ \mathrm{times} \label{n_plus}\\
\mathcal{H}_x&\underset{n\to n-1}{\longrightarrow}& -\alpha,\;  n\ \mathrm{times}\\
\mathcal{H}_x&\underset{n\to n+2}{\longrightarrow}& -1,\; (N-n-k) \ \mathrm{times}  \\
\mathcal{H}_x&\underset{n\to n-2}{\longrightarrow}& -1,  \; n-k \ \mathrm{times}
\end{eqnarray}
where $k$ is the number of groups of spins in the same direction. 

One gets 
\begin{eqnarray}
\mathcal{H}_z\mathcal{H}_x|n\rangle 
 &=& -2kE_n|n\rangle -(N-n-k)E_{n+2}|n+2\rangle-(n-k)E_{n-2}|n-2\rangle\nonumber\\
&-&\alpha (N-n)E_{n+1}|n+1\rangle-\alpha nE_{n-1}|n-1\rangle
\end{eqnarray}
Here $|n\rangle$ stays for a state with $n$ spins up. 

Finally the action  $\mathcal{H}_x$  will select the corresponding term and one obtains
\begin{eqnarray}
& &\langle \vec{n}|\mathcal{H}_x \mathcal{H}_z\mathcal{H}_x|\vec{n}\rangle = 
2kE_n  +(N-n-k)E_{n+2}+(n-k)E_{n-2}+\alpha^2 [(N-n)E_{n+1}+ nE_{n-1}]\nonumber\\
&=&(N-n)E_{n+2}+nE_{n-2}+\alpha^2 [(N-n)E_{n+1}+ nE_{n-1}]
= [N(1+a^2)-4-2a^2]E_n
\label{x_z_x}
\end{eqnarray}
Combining all terms together gives that for $N\geq 4$
\begin{eqnarray}
& &\mu_3(n)\equiv \langle \vec{n}| \mathcal{H}^3|\vec{n}  \rangle  = E_n^3+2 (N+k) E_n + (N-n-k)E_{n+2}+(n-k)E_{n-2}\\
&+& \alpha^2 [2E_n N+(N-n)E_{n+1}+nE_{n-1}-6N]=E_n^3+[3(1+\alpha^2)N-4-2\alpha^2]E_n-6N\alpha^2
 \nonumber
\end{eqnarray}
Notice that the dependence on $k$ disappears. 

Finally the third cumulant, $k_3=\mu_3-3\mu_2\mu_1+2\mu_1^3$, takes the following value
\begin{equation}
k_3=-6N\alpha^2-2E_n(\alpha^2+2)
\label{third}
\end{equation}
The next step is the calculation of the forth moment of Hamiltonian \eqref{h_x_h_z}
\begin{eqnarray}
& &\mathcal{H}^4= \mathcal{H}_x^4+\mathcal{H}_x^3\mathcal{H}_z+\mathcal{H}_x^2\mathcal{H}_z\mathcal{H}_x
+\mathcal{H}_x^2\mathcal{H}_z^2+\mathcal{H}_x\mathcal{H}_z\mathcal{H}_x^2
+\mathcal{H}_x\mathcal{H}_z\mathcal{H}_x\mathcal{H}_z+\mathcal{H}_x\mathcal{H}_z^2\mathcal{H}_x
+\mathcal{H}_x\mathcal{H}_z^3\nonumber\\
&+&\mathcal{H}_z \mathcal{H}_x^3+\mathcal{H}_z \mathcal{H}_x^2 \mathcal{H}_z+\mathcal{H}_z \mathcal{H}_x \mathcal{H}_z\mathcal{H}_x+\mathcal{H}_z \mathcal{H}_x\mathcal{H}_z^2
+\mathcal{H}_z^2\mathcal{H}_x^2+\mathcal{H}_z^2\mathcal{H}_x\mathcal{H}_z+\mathcal{H}_z^3\mathcal{H}_x+\mathcal{H}_z^4 
\end{eqnarray}
When calculating $\langle \vec{n} |\mathcal{H}^4|\vec{n}\rangle $  many terms are known from the above expressions
\begin{eqnarray}
& &
\langle \vec{n} | \mathcal{H}^4|\vec{n}\rangle =
\langle \vec{n} | \mathcal{H}_x^4|\vec{n}\rangle +
2 \langle \vec{n} | \mathcal{H}_x^3 |\vec{n}\rangle E_n+
3\langle \vec{n} | \mathcal{H}_x^2 |\vec{n}\rangle E_n^2+
4\langle \vec{n} |\mathcal{H}_x |\vec{n}\rangle E_n^3  \\
&+&
2\langle \vec{n} |\mathcal{H}_x\mathcal{H}_z\mathcal{H}_x|\vec{n}\rangle E_n
 + E_n^4 + 
\langle \vec{n} | \mathcal{H}_x^2\mathcal{H}_z\mathcal{H}_x|\vec{n}\rangle+
\langle \vec{n} | \mathcal{H}_x\mathcal{H}_z\mathcal{H}_x^2|\vec{n}\rangle + 
\langle \vec{n} | \mathcal{H}_x\mathcal{H}_z^2\mathcal{H}_x |\vec{n}\rangle 
\nonumber
\end{eqnarray}
Only three last quantities require separate calculation.

Term $\langle \vec{n} |\mathcal{H}_x\mathcal{H}_z^2\mathcal{H}_x|\vec{n}\rangle$ can be performed  as above and the result is the same as Eq.~\eqref{x_z_x} but with substitution $E_m\to E_m^2$
\begin{eqnarray}
& &\langle \vec{n}|\mathcal{H}_x \mathcal{H}_z^2\mathcal{H}_x|\vec{n}\rangle = 
2kE_n^2  +(N-n-k)E_{n+2}^2+(n-k)E_{n-2}^2+\alpha^2 (N-n)E_{n+1}^2+\alpha^2 nE_{n-1}^2\nonumber\\
&=& -32k\lambda^2 +(N-8)E_n^2+16N\lambda^2 +\alpha^2[(N-4)E_n^2+4N\lambda^2]
\label{x_z_z_x}
\end{eqnarray}
Operator $\mathcal{H}_x$ flips one or two nearby spins, therefore only the following contribution are non-zero
\begin{eqnarray}
& &\langle \vec{n} | \mathcal{H}_x^2\mathcal{H}_z\mathcal{H}_x|\vec{n}\rangle=-
2\alpha^2 \langle \vec{n} |\Big (\sum_n\sigma_n^x\sigma_{n+1}^x\Big )\ \Big (\sum_{n_1} \sigma_{n_1}^x\Big ) \mathcal{H}_z \Big ( \sum_{n_2} \sigma_{n_2}^x\Big )|\vec{n}\rangle \nonumber \\
&-& \alpha^2 \langle \vec{n} |\Big ( \sum_{n_1} \sigma_{n_1}^x \Big )\ \Big ( \sum_{n_2} \sigma_{n_2}^x \Big )\mathcal{H}_z \Big ( \sum_n\sigma_n^x\sigma_{n+1}^x\Big )|\vec{n}\rangle
\end{eqnarray}
Consider the first term. Operator $\sum_{n_2} \sigma_{n_2}^x$ acting on $|\vec{n}\rangle$ produces $N-n$ states with $n+1$ spins up and $n$ states with $n-1$ spins down (cf. \eqref{n_plus} and \eqref{n_minus}). Operator $\mathcal{H}_z $ multiplies them by $E_{n+1}$ and $E_{n-1}$ correspondingly. Other terms should combine to produce the initial state. In the end one gets
\begin{eqnarray}
& &\langle \vec{n} | \mathcal{H}_x^2\mathcal{H}_z\mathcal{H}_x|\vec{n}\rangle
=-4\alpha^2 [(N-n)E_{n+1}+nE_{n-1} ]\\ 
&-& 2\alpha^2[2kE_n+(N-n-k)E_{n+2}+(n-k)E_{n-2} ]
= -2\alpha^2(3N-8)E_n\nonumber 
\end{eqnarray}
Taking into account all terms we obtain that for $N\geq 5$ 
\begin{eqnarray}
\mu_4(n,k)&\equiv & \langle \vec{n} | \mathcal{H}^4|\vec{n}\rangle=
E_n^4+E_n^2[6N(1+\alpha^2)-16-8\alpha^2]+8 E_n(4-3N)\alpha^2+3N^2(1+\alpha^2)^2\nonumber \\
&-& 32k\lambda^2+N(24\alpha^2-2-2\alpha^4+16\lambda^2+4\lambda^2\alpha^2)
\label{forth} 
\end{eqnarray}

The forth cumulant by definition is the following combination, $k_4\equiv\mu_4-4\mu_3\mu_1-3\mu_2^2+12\mu_2\mu_1^2-6\mu_1^4 $. Simple calculations  gives
\begin{equation}
k_4=32\alpha^2 E_n-32k\lambda^2+N(24\alpha^2-2-2\alpha^4+16\lambda^2+4\lambda^2\alpha^2)
\end{equation} 
In the calculations we used instead of $k$ its mean value of $k$ in all states with fixed $n$, $\langle k\rangle=n(N-n)/(N-1)$ \cite{a_b}.

All above calculations were performed without the selecting a particular translational momentum. If instead of basis set 
$|\vec{n}\rangle$ one uses states which are eigenstates of the translational operator as in \eqref{translational_states}, moments may be different. We check that for lower order moments the difference is small and does not change noticeable the results. For simplicity we ignore such corrections.  


\end{document}